\documentclass[a4paper,11pt]{article}

\usepackage{jheppub} 

\usepackage[T1]{fontenc} 
\usepackage{mathrsfs}

\title{\boldmath   Covariant Chiral Kinetic Equation in Non-Abelian Gauge field from ``covariant gradient expansion''}

\author[a]{Xiao-Li Luo,}
\author[a,1]{Jian-Hua Gao.\note{Corresponding author.}}

\affiliation[a]{Shandong Provincial Key Laboratory of Optical Astronomy and Solar-Terrestrial
Environment, Institute of Space Sciences, Shandong University, Weihai,
Shandong 264209, China}

\emailAdd{xiaoli\_luo@mail.sdu.edu.cn}
\emailAdd{gaojh@sdu.edu.cn}

\abstract{We derive the chiral kinetic equation in 8 dimensional phase space in  non-Abelian  $SU(N)$ gauge field  within
the Wigner function formalism.  By using the ``covariant gradient expansion'', we disentangle the Wigner equations in four-vector space up to the first order
and find that only the time-like  component of the chiral Wigner function is independent while other components can be explicit derivative. After further decomposing
the Wigner function or equations in color space, we present the non-Abelian  covariant chiral kinetic equation
for the color singlet and multiplet  phase-space distribution functions.
These  phase-space distribution functions have non-trivial Lorentz transformation rules when we define them in different reference frames.
The chiral anomaly from non-Abelian gauge field arises naturally  from the Berry monopole in Euclidian momentum space in the vacuum or Dirac sea contribution.
The anomalous currents as non-Abelian counterparts of chiral magnetic effect and chiral vortical effect have also been derived from the non-Abelian chiral kinetic equation. }

\begin{document}
\maketitle
\flushbottom

\section{Introduction}
\label{sec:intro}

   In recent years, there has been a considerable amount of theoretical work on the chiral kinetic theory (CKT) in relativistic heavy ion collisions.  The  CKT aims to  incorporate the chiral anomaly
into  kinetic theory  and provide a consistent formalism to describe various novel chiral effects, e.g., chiral magnetic effect~\cite{Vilenkin:1980fu,Kharzeev:2007jp,Fukushima:2008xe},
chiral vortical effect~\cite{Vilenkin:1978hb,Kharzeev:2007tn,Erdmenger:2008rm,Banerjee:2008th}, chiral separation effect~\cite{Son:2004tq,Metlitski:2005pr}  and so on,
which are all associated with the chiral anomaly. Recent progress on chiral effects and chiral kinetic theory in relativistic heavy ion collisions can be found in the reviews such as
\cite{Kharzeev:2013ffa,Kharzeev:2015znc,Liu:2020ymh,Gao:2020vbh,Gao:2020pfu}.  The chiral kinetic equation has been derived from various methods, such as  semiclassical approach
\cite{Duval:2005vn,Wong:2011nt,Son:2012wh,Stephanov:2012ki,Stone:2013sga,Dwivedi:2013dea,Akamatsu:2014yza,Chen:2014cla,Manuel:2014dza,Hayata:2017ihy},
Wigner function formalism \cite{Gao:2012ix,Chen:2012ca,Hidaka:2016yjf,Huang:2018wdl,Gao:2018wmr,Liu:2018xip},
effective field theory \cite{Son:2012zy,Carignano:2018gqt,Lin:2019ytz,Carignano:2019zsh} and world-line approach
\cite{Mueller:2017lzw,Mueller:2017arw,Mueller:2019gjj}.  The  numerical simulation based on chiral kinetic equation can be found  in Refs.
\cite{Sun:2016nig,Sun:2016mvh,Sun:2017xhx,Sun:2018idn,Sun:2018bjl,Zhou:2018rkh,Zhou:2019jag,Liu:2019krs} .

Despite all these development, so far most of the literature focuses  on the CKT in Abelian gauge field. Only very restricted
work \cite{Stone:2013sga,Akamatsu:2014yza,Hayata:2017ihy,Mueller:2019gjj} had discussed the CKT in non-Abelian gauge field.
However, as we all know, the dynamics of the produced quark-gluon plasma in relativistic heavy ion collisions are mainly
determined by quantum chromodynamics ---  non-Abelian $SU(3)$ gauge field.  Especially, in the small $x$ physics, the initial
state in relativistic nucleus-nucleus collisions can be described as a classical coherent non-Abelian gauge field
configuration called the color glass condensate\cite{Gribov:1984tu,Mueller:1985wy,McLerran:1993ka,McLerran:1994vd,Iancu:2003xm}.
It still remains an open question how the decoherence from the classical color
field to the quark gluon plasma takes place.  In order to address these problems, we need generalize the  CKT in Abelian gauge field to
the one in non-Abelian gauge field.

In this paper, we will be dedicated to deriving the chiral kinetic equation in  $SU(N)$ gauge field from the quantum transport theory
\cite{Heinz:1983nx,Elze:1986hq,Elze:1986qd,Elze:1989un,Ochs:1998qj} based on the Wigner functions from  quantum gauge field theory.
In Sec.\ref{sec:quantum}, we review the Wigner function formalism given in Refs.  \cite{Heinz:1983nx,Elze:1986hq,Elze:1986qd,Elze:1989un}
and present some results  in  Ref.\cite{Ochs:1998qj} that
would be useful for our present work. In Sec.\ref{sec:disentangling}, we apply the ``covariant gradient expansion'' given in \cite{Elze:1986hq,Elze:1986qd,Elze:1989un,Ochs:1998qj}
to expanding the Wigner equations for massless fermions
 up to the first order and  disentangle the Wigner equations
by  the method developed in the Abelian case in Ref.\cite{Gao:2018wmr} .
We find that only the timelike  component of the Wigner functions is independent
and all other spacelike components can be derivative from timelike component directly. Such result is very similar to the Abelian case and  reduces the Wigner equations greatly.
We present the covariant chiral kinetic equation for this independent Wigner function in 8-dimensional form, i.e.,
4-dimensional momentum space and 4-dimensional coordinate space. In comparison with the Abelian case, the extra constraint equation appears in non-Abelian case.
 In Sec.\ref{sec:color}, we decompose the results further in the color space and find that the color singlet  phase-space distribution function and
multiplet ones are totally coupled with each other. In Sec.\ref{sec:Lorentz},
We  discuss the modified  Lorentz transformation  of the distribution function in phase space when we define it in different reference frames.
With the results in previous sections, we calculate the vector and axial currents induced by color field and vorticity in Sec.\ref{sec:currents}.
It turns out that the non-Abelian chiral anomaly can be derived directly from the 4-dimentional Berry curvature in the vacuum contribution of the color singlet Wigner function. With specific distribution near global equilibrium, we can obtain the non-Abelian counterparts of chiral magnetic effect and chiral vortical effect.  Finally, we summarize the paper in Sec.\ref{sec:summary}.

In this work, we use the convention for the metric $g^{\mu\nu}=\mathrm{diag}(1,-1,-1,-1)$,
Levi-Civita tensor $\epsilon^{0123}=1$. We choose natural units such that $\hbar=c=1$  except for the cases
 when we want to display  $\hbar$ dependence to  clarify the
perturbative expansion.

\section{Quantum transport theory}
\label{sec:quantum}

In quantum transport theory, the gauge invariant density matrix for  spin-1/2 quarks is defined  as \cite{Heinz:1983nx,Elze:1986hq,Elze:1986qd}
\begin{eqnarray}
\label{density}
\rho \left(x+\frac{y}{2},x-\frac{y}{2}\right) = \bar\psi\left(x+\frac{y}{2}\right)
U\left(x+\frac{y}{2},x\right)\otimes U\left(x,x-\frac{y}{2}\right) \psi\left(x-\frac{y}{2}\right).
\end{eqnarray}
where the direct product  is over both spinor and color indices. The element of  density matrix with specific color and spinor indices is given by
\begin{eqnarray}
\label{density-element}
\rho_{\alpha\beta}^{ij} \left(x+\frac{y}{2},x-\frac{y}{2}\right) = \bar\psi_\beta^{j'}\left(x+\frac{y}{2}\right)
U^{j'j}\left(x+\frac{y}{2},x\right) U^{ii'}\left(x,x-\frac{y}{2}\right) \psi^{i'}_{\alpha}\left(x-\frac{y}{2}\right).
\end{eqnarray}
where $\alpha,\beta$ denote spinor indices, $i,i',j,j'$ mean color indices in fundamental representation and $U^{j'j}$ or $U^{i'i}$
is the  Wilson line  or gauge link
\begin{eqnarray}
\label{link}
U^{ij}(x,y)=\left[{P}\exp\left(\frac{ig}{\hbar  }\int_y^x dz^\mu  A_\mu(z)\right)\right]^{ij}
\end{eqnarray}
which is necessary to   keep the  operator gauge invariant. In the definition of Wilson line  $P$ denotes  path ordering of the operator and the integral in the exponent is taken along the straight path
from $x$ to $y$. The gauge field potential is defined by $A_\mu=A^a_\mu t^a$, with the $N^2-1$ hermitian generators of $SU(N)$ in the fundamental representation satisfying
\begin{eqnarray}
\label{Lie}
\textrm{Tr}\, t^a =0,\ \ \
\left[t^a,t^b\right]= if^{abc}t^c,\ \ \
\left\{t^a,t^b\right\} = \frac{1}{N}\delta^{ab}{\bf 1}+ d^{abc}t^c.
\end{eqnarray}
For  non-Abelian gauge field,  the covariant derivative in the fundamental representation is defined as,
\begin{eqnarray}
\label{Dmu}
D_\mu(x)&=&\partial_\mu -\frac{ig}{\hbar} A_\mu(x),
\end{eqnarray}
and the field strength tensor follows as
\begin{eqnarray}
\label{Fmunu}
F_{\mu\nu}(x)&\equiv& F_{\mu\nu}^a t^a =-\frac{\hbar }{ig} \left[ D_\mu,D_\nu\right] = \partial_\mu A_\nu(x)-\partial_\nu A_\mu(x)-\frac{ig}{\hbar }\left[A_\mu(x),A_\nu(x)\right].
\end{eqnarray}
The Wigner operator  $\hat W(x,p)$ is related to  the gauge invariant density matrix  by  Fourier transformation
\begin{eqnarray}
\label{wigner-hat}
\hat W(x,p)=\int\frac{d^4 y}{(2\pi)^4} e^{-ip\cdot y}  \rho \left(x+\frac{y}{2},x-\frac{y}{2}\right),
\end{eqnarray}
and the Wigner function is  defined as ensemble averaging of the Wigner operator
\begin{eqnarray}
\label{wigner}
W(x,p)=\langle \hat W(x,p) \rangle.
\end{eqnarray}
 In our present work, we will concentrate on the quark matter under a purely classical external non-Abelian gauge field, in which ordinary matrix multiplication  rules in spinor space or color space
suffice and the Wigner equations will not generate the so-called BBGKY-hierarchy \cite{Groot:1980} and can be closed by itself
\begin{eqnarray}
\label{equation-5}
& &\left[m-\gamma^\mu \left(p_\mu +\frac{1}{2}i\mathscr{D}_\mu(x)\right)\right] W(x,p)\nonumber\\
&=&\frac{i g}{2}\gamma^\mu \partial^\nu_p \left\{  \int_0^1 d s  {\,} \frac{1+s}{2}\left[ e^{-\frac{1}{2} i s\Delta}  F_{\mu\nu}(x)\right]  W(x,p)\right.\nonumber\\
& &\left. \hspace{1.5cm}+  W(x,p)\int_0^1 d s  {\,} \frac{1-s}{2}
\left[ e^{\frac{1}{2}is\Delta} F_{\mu\nu}(x) \right]\right\},
\end{eqnarray}
together with the hermitian adjoint equation
\begin{eqnarray}
\label{equation-6}
& &  W(x,p)\left[m-\gamma^\mu \left(p_\mu -\frac{1}{2}i\mathscr{D}^\dagger_\mu(x)\right)\right]\nonumber\\
&=&-\frac{i g}{2} \partial^\nu_p \left\{  \int_0^1 d s  {\,} \frac{1-s}{2}
\left[ e^{-\frac{1}{2}is\Delta} F_{\mu\nu}(x) \right]  W(x,p)\right.\nonumber\\
& &\left.\hspace{1.5cm} +  W(x,p) \int_0^1 d s  {\,} \frac{1+s}{2}\left[ e^{\frac{1}{2} i s\Delta}  F_{\mu\nu}(x)\right]\right\}
\gamma^\mu,
\end{eqnarray}
where we have introduced the definition of  covariant derivative in the adjoint representation for a second-rank tensor $\mathcal{T}(x)$ in color space by
\begin{eqnarray}
\mathscr{D}_\mu (x) \mathcal{T}(x) \equiv \left[{D}_\mu (x), \mathcal{T}(x) \right]=\partial_\mu^x \mathcal{T}(x) - \frac{i g}{\hbar} \left[ A_\mu(x), \mathcal{T}(x)\right],
\end{eqnarray}
and $\Delta\equiv \partial_p \cdot \mathscr{D}(x)$ with  $\mathscr{D}(x)$ only acting on $ F_{\mu\nu}$ and $\partial_p$ always on $ W$ after or in front of it.
It should be noted  that in the definition of the Wigner function given by Eq.~(\ref{wigner})
and the Wigner equations  (\ref{equation-5}) and (\ref{equation-6})
there is no normal ordering in the Wigner matrix because we did not make any manipulation on the order of the quark field.
It has been demonstrated in \cite{Gao:2019zhk,Fang:2020com} that this  plays a central role to give rise to the chiral anomaly in the quantum kinetic theory.

If we  take the convention in \cite{Ochs:1998qj}, momentum derivatives
standing to the right of the Wigner function are defined in the sense of partial integration as
\begin{eqnarray}
\label{convention}
  W(x,p) \partial^{\nu_1}_p \cdots  \partial^{\nu_k}_p\equiv (-1)^k  \partial^{\nu_k}_p \cdots  \partial^{\nu_1}_p  W(x,p),
\end{eqnarray}
and define generalized non-local momentum and derivative operators  $\Pi_\mu$
and $G_\mu$ as
\begin{eqnarray}
\label{Pi}
\Pi_\mu &=& p_\mu  + \frac{g}{2}\int_0^1 d s  {\,} \left( e^{-\frac{1}{2} i s\Delta}  F_{\mu\nu}(x)\right)  i s \partial^\nu_p, \nonumber\\
G_\mu &=& D_\mu  + \frac{g}{2}\int_0^1 d s  {\,} \left( e^{-\frac{1}{2} i s\Delta}  F_{\mu\nu}(x)\right)    \partial^\nu_p,
\end{eqnarray}
 the Wigner equations  can be cast into a more compact form \cite{Ochs:1998qj},
\begin{eqnarray}
\label{equation-L-1}
2m   W(x,p) &=& \gamma^\mu \left(\left\{\Pi_\mu,   W(x,p)\right\} + i \left[G_\mu,   W(x,p)\right]\right),\\
\label{equation-R-1}
2m   W(x,p) &=&  \left(\left\{\Pi_\mu,   W(x,p)\right\} - i \left[G_\mu,   W(x,p)\right]\right)\gamma^\mu.
\end{eqnarray}
Adding or subtracting the two equations above gives
\begin{eqnarray}
\label{equation-LR1}
 4m   W(x,p) &=& \left\{\gamma^\mu,\left\{\Pi_\mu,   W(x,p)\right\}\right\}
 + i \left[ \gamma^\mu, \left[G_\mu,   W(x,p)\right]\right],\\
\label{equation-LR2}
0 &=& \left[\gamma^\mu,\left\{\Pi_\mu,   W(x,p)\right\}\right]
 + i \left\{ \gamma^\mu, \left[G_\mu,   W(x,p)\right]\right\}.
\end{eqnarray}
In spinor space, we can decompose the Wigner function into
\begin{eqnarray}
\label{decomposition}
  W=\frac{1}{4}\left[{ {\mathscr{F}}}+i\gamma^5{ {\mathscr{P}}}+\gamma^\mu { { \mathscr{V}}}_\mu +\gamma^\mu\gamma^5{  {\mathscr{A}}}_\mu
+\frac{1}{2}\sigma^{\mu\nu} { {\mathscr{S}}}_{\mu\nu}\right].
\end{eqnarray}
In this work, we will restrict ourselves to the  massless or chiral fermions. In consequence, if we introduce a chirality basis via
\begin{eqnarray}
\label{Js}
 {\mathscr{J}}^\mu_s &=& \frac{1}{2}\left( {\mathscr{V}}^\mu + s {\mathscr{A}}^\mu\right),
\end{eqnarray}
where $s =+1/-1$  denotes right-handed/left-handed component, the equations for the chiral Wigner function ${\mathscr{J}}^\mu_s$ will decouple from all the other components of the
Wigner function and each other as well,
which leads to
\begin{eqnarray}
\label{Js-c1}
0 &=& \left\{\Pi_\mu,  {\mathscr{J}}^\mu_s \right\},\\
\label{Js-t}
0 &=& \left[G_\mu,  {\mathscr{J}}^\mu_s \right],\\
\label{Js-c2}
0&=& \left\{\Pi^\mu,  {\mathscr{J}}^{\nu}_s \right\} - \left\{\Pi^\nu,  {\mathscr{J}}^{\mu}_s \right\}
 {+} s \hbar \epsilon^{\mu\nu\alpha\beta} \left[G_\alpha,  {\mathscr{J}}_{s\beta} \right],
\end{eqnarray}
where we have recovered the $\hbar$ dependence before the generalized derivative operators  in the last equation in order to make perturbative expansion in the following section.
These Wigner equations will be  the starting point of our present work in the following.
For brevity, we will suppress the subscript $s$ of the left-hand or right-hand Wigner function ${\mathscr{J}}^\mu_s$ in the subsequent sections
and reinstate it when it is necessary.

\section{Disentangling  Wigner equations in four-vector space}
\label{sec:disentangling}

In  the Abelian plasma,  the disentanglement  theorem of Wigner functions has been demonstrated in Ref. \cite{Gao:2018wmr}, which tell us that
up to any order of $\hbar$ among four components of Wigner functions $\mathscr{J}^\mu$ only the timelike component
is independent and satisfies only one independent Wigner equation,  the other spatial components can be totally fixed from this independent Wigner function
 and the Wigner equations for them are all satisfied automatically.  Now let us try to generalize this disentanglement formalism   from Abelian gauge field to non-Abelian gauge field.
In order to achieve this goal, we will resort to the ``covariant gradient expansion'' proposed in Refs. \cite{Elze:1986qd,Elze:1989un,Ochs:1998qj} . In this expansion scheme,
when we have one extra covariant derivative $D_\mu$ or $\mathscr{D}_\mu$, we will have one extra higher order contribution. The ``covariant gradient expansion''  preserves
gauge invariance order by order automatically. Actually we can trace such expansion in powers of $\hbar$, e.g., in the Wigner equations (\ref{Js-c2}) and the generalized
non-local momentum and derivative operators
\begin{eqnarray}
\label{Pi-1}
\Pi_\mu &=& \sum_{k=0}^\infty \hbar^k \Pi_\mu^{(k)} = p_\mu  -\hbar \frac{i g}{2 }\sum_{k=0}^\infty \left(-\frac{i\hbar}{2}\right)^k  \frac{k+1}{(k+2)!}
\left[\left(\partial_p\cdot  {\mathscr{D}} \right)^k F_{\nu\mu}\right] \partial_p^\nu \\
\label{G-1}
G_\mu &=&\sum_{k=0}^\infty \hbar^{k} G_\mu^{(k)} = D_\mu  - \frac{ g}{2 }\sum_{k=0}^\infty \left(-\frac{i\hbar}{2}\right)^k  \frac{1}{(k+1)!}
\left[\left(\partial_p\cdot  {\mathscr{D}} \right)^k F_{\nu\mu} \right] \partial_p^\nu.
\end{eqnarray}
Up to the second order of $\hbar$, the non-local operators $\Pi_\mu$ and $G_\mu$ are given by
\begin{eqnarray}
\label{Pi-012}
\Pi_\mu^{(0)} &=& p_\mu,\ \ \
\Pi_\mu^{(1)} = \frac{i g}{4} F_{\mu\nu}\partial^\nu_p,\ \
\Pi_\mu^{(2)} = \frac{ g}{12 }\left[\left(\partial_p\cdot  {\mathscr{D}} \right) F_{\mu\nu} \right] \partial_p^\nu,\\
\label{G-01}
G_\mu^{(0)} &=&  D_\mu  +\frac{g}{2} F_{\mu\nu}\partial^\nu_p, \ \
G_\mu^{(1)} = - \frac{i g}{8}\left[\left(\partial_p\cdot  {\mathscr{D}} \right) F_{\mu\nu} \right] \partial_p^\nu.
\end{eqnarray}
We can also expand the Wigner operator as
\begin{eqnarray}
  W (x,p) =\sum_{k=0}^\infty \hbar^k   W^{(k)}(x,p).
\end{eqnarray}
However it should be noted that the ``covariant gradient expansion''  is not completely identical to an expansion in powers of $\hbar$ for non-Abelian gauge field which had been pointed out in
\cite{Elze:1986qd,Elze:1989un,Ochs:1998qj}  though it is identical for Abelian gauge field.
In non-Abelian case, there is an extra gauge potential $A_\mu$ with $ig/\hbar$ in the covariant derivative $D_\mu$ or $\mathscr{D}_\mu$  in Eqs. (\ref{Pi-1}) and (\ref{G-1}) while there only exist ordinary
derivative $\partial_\mu^x$ in the Abelian case.

In order to disentangle the Wigner equations further, it is convenient to introduce  time-like  4-vector $n^\mu$ with normalization $n^2=1$.
For simplicity we assume $n^\mu$ is a constant vector. With the auxiliary vector $n^\mu$,  we can  decompose any vector $X^\mu$ into
the component parallel to $n^\mu$ and the other components perpendicular to $n^\mu$,
\begin{equation}
X^\mu=X_n n^\mu + \bar X^\mu,
\end{equation}
where $X_n=X\cdot n$ and $\bar X^\mu = \Delta^{\mu\nu}X_\nu$ with $\Delta^{\mu\nu}=g^{\mu\nu}-n^\mu n^\nu$.
The gauge field tensor $F^{\mu\nu}$ can be also decomposed into
\begin{equation}
F^{\mu\nu}=E^\mu n^\nu -E^\nu n^\mu -\bar\epsilon^{\mu\nu\sigma} B_\sigma
\end{equation}
with
\begin{equation}
E^\mu=F^{\mu\nu}n_\nu ,\ \ B^\mu=\frac{1}{2}\bar\epsilon^{\mu\rho\sigma} F_{\rho\sigma},
\end{equation}
where for notational convenience we have defined $\bar\epsilon^{\mu\alpha\beta}=\epsilon^{\mu\nu\alpha\beta}n_\nu$.

Now we can decompose the Wigner functions and Wigner equations  along the direction $n^\mu$ order by order.
The leading order or the zeroth order result is very simple
\begin{eqnarray}
\label{J-c1-n-0-W}
0 &=& p_n {\mathscr{J}}_n^{(0)} +\bar p_\mu \bar{{\mathscr{J}}}^{(0)\mu},\\
\label{J-t-n-0-W}
0 &=& \left[G_n^{(0)},{\mathscr{J}}_n^{(0)}\right]+\left[\bar G_\mu^{(0)}, \bar{{\mathscr{J}}}^{(0)\mu} \right],\\
\label{J-c2-n-0-W}
0&=& \bar p ^\mu {\mathscr{J}}_n^{(0)}  - p_n \bar{{\mathscr{J}}}^{(0)\mu},\\
\label{J-c2-bar-0-W}
0&=& \bar p^\mu \bar{{\mathscr{J}}}^{(0)\nu} - \bar p^\nu \bar{ {\mathscr{J}}}^{(0)\mu}.
\end{eqnarray}
From Eq.(\ref{J-c2-n-0-W}), we can express the space-like component $\bar{{\mathscr{J}}}^{(0)\mu}$ in terms of ${\mathscr{J}}_n^{(0)}$
\begin{eqnarray}
\label{Jnbar-Jn-0}
\bar{{\mathscr{J}}}^{(0)\mu} &=& \bar p ^\mu   \frac{{\mathscr{J}}_n^{(0)} }{p_n}.
\end{eqnarray}
Substituting this relation into Eqs.(\ref{J-c1-n-0-W}) gives rise to the on-shell condition
\begin{eqnarray}
\label{Jn0-onshell}
p^2 \frac{{\mathscr{J}}_n^{(0)}}{p_n} &=& 0,
\end{eqnarray}
which means ${{\mathscr{J}}_n^{(0)}}/{p_n}$ must be proportional to the Dirac delta function $\delta(p^2)$
\begin{eqnarray}
\label{Jn-0}
\frac{{\mathscr{J}}_n^{(0)}}{p_n} &=& f^{(0)}\delta(p^2),
\end{eqnarray}
where $f^{(0)}$ can be regarded as the usual particle distribution function in four-dimensional momentum space and
four-dimensional coordinate space. It must be non-singular function at $p^2=0$. Putting Eqs.(\ref{Jn-0}) and (\ref{Jnbar-Jn-0}) together, we get
the full Wigner function of the zeroth order
\begin{eqnarray}
\label{Jmu-f-0}
{\mathscr{J}}^{(0)\mu} &=& p ^\mu    f^{(0)}\delta(p^2).
\end{eqnarray}
The transport equation satisfied by $f^{(0)}$ can be obtained from Eq.(\ref{J-t-n-0-W})
\begin{eqnarray}
\label{f0-evolve}
0&=&\left[ G_\mu^{(0)}, p^\mu  f^{(0)}\delta(p^2) \right] .
\end{eqnarray}
It is obvious that Eq.(\ref{J-c2-bar-0-W}) is automatically satisfied with the expression (\ref{Jnbar-Jn-0}).

The next-to-leading order or the first order equations are given by
\begin{eqnarray}
\label{J-c1-n-1-W}
0 &=& 2 p_n{\mathscr{J}}^{(1)}_n +2  \bar p_\mu \bar{{\mathscr{J}}}^{(1)\mu}
+\left\{\Pi_n^{(1)}, {\mathscr{J}}^{(0)}_n \right\} + \left\{\bar\Pi_\mu^{(1)}, \bar{{\mathscr{J}}}^{(0)\mu} \right\},\\
\label{J-t-n-1-W}
0 &=& \left[G_n^{(0)}, {\mathscr{J}}^{{(1)}}_n \right]+\left[\bar G_\mu^{(0)}, \bar{{\mathscr{J}}}^{{(1)}\mu} \right]
+\left[G_n^{(1)}, {\mathscr{J}}^{{(0)}}_n \right]+\left[\bar G_\mu^{(1)},\bar{ {\mathscr{J}}}^{{(0)}\mu} \right],\hspace{8pt}\\
\label{J-c2-n-1-W}
0 &=& 2\bar p^\mu {\mathscr{J}}^{(1)}_n - 2 p_n \bar{{\mathscr{J}}}^{(1)\mu}
 + \left\{\bar\Pi^{(1)\mu},{\mathscr{J}}^{(0)}_n \right\}   -\left\{\Pi^{(1)}_n, \bar{{\mathscr{J}}}^{(0)\mu} \right\}\nonumber\\
& &  {+}s \bar\epsilon^{\mu\alpha\beta} \left[G_\alpha^{(0)}, {\mathscr{J}}_\beta^{(0)} \right],\\
 \label{J-c2-bar-1-W}
0 &=&2\bar p^{\mu} \bar{{\mathscr{J}}}^{(1)\nu}
 -2\bar p^{\nu} \bar{{\mathscr{J}}}^{(1)\mu}
 + \left\{\bar\Pi^{(1)\mu},\bar{ {\mathscr{J}}}^{(0)\nu} \right\}
 -\left\{\bar\Pi^{(1)\nu}, \bar{{\mathscr{J}}}^{(0)\mu} \right\}\nonumber\\
& & {+} s\bar \epsilon^{\mu\nu\alpha}\left( \left[G_\alpha^{(0)},{\mathscr{J}}_n^{(0)} \right]
-\left[G_n^{(0)}, {\mathscr{J}}_\alpha^{(0)} \right]\right).
\end{eqnarray}
From Eq.(\ref{J-c2-n-1-W}), we can express $\bar{{\mathscr{J}}}^{(1)\mu}$ in terms of ${\mathscr{J}}^{(1)}_n$ and ${\mathscr{J}}^{(0)}_n$
\begin{eqnarray}
\label{Jnbar-Jn-1a}
\bar{{\mathscr{J}}}^{(1)\mu} &=&\bar p^\mu\frac{ {\mathscr{J}}^{(1)}_n}{p_n}
 {+}\frac{s}{2 p_n}\bar\epsilon^{\mu\alpha\beta} \left[G_\alpha^{(0)}, \bar p_\beta\frac{ {\mathscr{J}}_n^{(0)}}{p_n} \right]\nonumber\\
& &+ \frac{1}{2 p_n}\left(\left\{\bar\Pi^{(1)\mu}, p_n \frac{{\mathscr{J}}^{(0)}_n}{p_n} \right\}
-  \left\{\Pi^{(1)}_n,  \bar p^\mu \frac{{\mathscr{J}}_n^{(0)}}{p_n} \right\}\right).
\end{eqnarray}
Substituting it into Eqs.(\ref{J-c1-n-1-W}) and (\ref{J-t-n-1-W}) gives rise to the modified on-shell condition and transport equation for ${\mathscr{J}}_n^{(1)}$, respectively,
\begin{eqnarray}
\label{Jn1-onshell}
p^2 \frac{{\mathscr{J}}_n^{(1)}}{p_n} &=& {-} \frac{s}{2 p_n}\bar\epsilon^{\mu\alpha\beta} \bar p_\mu \left[G_\alpha^{(0)}, \bar p_\beta\frac{ {\mathscr{J}}_n^{(0)}}{p_n} \right]
-{\frac{1}{2}} \left\{\Pi_\mu^{(1)},p^\mu \frac{{\mathscr{J}}^{(0)}_n}{p_n} \right\}\nonumber\\
& & - \frac{\bar p_\mu}{2 p_n}\left(\left\{\bar\Pi^{(1)\mu}, p_n \frac{{\mathscr{J}}^{(0)}_n}{p_n} \right\}
-  \left\{\Pi^{(1)}_n,  \bar p^\mu \frac{{\mathscr{J}}_n^{(0)}}{p_n} \right\}\right),\\
\label{Jn1-evolve}
\left[ G_\mu^{(0)}, p^\mu \frac{{\mathscr{J}}^{(1)}_n}{p_n} \right] &=&
 {-}\frac{s}{2}\bar\epsilon^{\mu\alpha\beta}\left[\bar G_\mu^{(0)},
\frac{1}{p_n} \left[G_\alpha^{(0)}, \bar p_\beta\frac{ {\mathscr{J}}_n^{(0)}}{p_n} \right]\right]
-\left[ G_\mu^{(1)}, p^\mu \frac{ {\mathscr{J}}^{{(0)}}_n }{p_n}\right]\nonumber\\
& & -\frac{1}{2}\left[\bar G_\mu^{(0)}, \frac{1}{ p_n}\left(\left\{\bar\Pi^{(1)\mu}, p_n \frac{{\mathscr{J}}^{(0)}_n}{p_n} \right\}
-  \left\{\Pi^{(1)}_n,  \bar p^\mu \frac{{\mathscr{J}}_n^{(0)}}{p_n} \right\}\right) \right].
\end{eqnarray}
It is easy to verify that the general expression of the constraint equation (\ref{Jn1-onshell}) is given by
\begin{eqnarray}
\label{Jn1-onshell-a}
\frac{{\mathscr{J}}_n^{(1)}}{p_n} = f^{(1)}\delta(p^2)  {+} \frac{s}{2 p_n}\bar\epsilon^{\mu\alpha\beta}  p_\mu
\left\{\frac{g}{2}F_{\alpha\beta},f^{(0)} \right\}\delta'(p^2)
+ \left\{\Pi_\mu^{(1)},p^\mu f^{(0)} \right\} \delta'(p^2).
\end{eqnarray}
Just like $ f^{(0)}$, the function $ f^{(1)}$ is also a non-singular   distribution function  at $p^2=0$ in four-dimensional momentum space and
four-dimensional coordinate space and   can be regarded as the first order correction to $ f^{(0)}$. The transport equation for $ f^{(1)}$ can be directly obtained by
inserting Eq.(\ref{Jn1-onshell-a}) into Eq.(\ref{Jn1-evolve}) and will not be presented explicitly here to avoid too lengthy equations.
Putting Eqs.(\ref{Jnbar-Jn-1a}) and (\ref{Jn1-onshell-a}) together, we get
the full Wigner function of the first order
\begin{eqnarray}
\label{Jmu-f-1}
{\mathscr{J}}^{(1)\mu} &=& p^\mu\left[f^{(1)}\delta(p^2)  {+} \frac{s}{2 p_n}\bar\epsilon^{\nu\alpha\beta}  p_\nu
\left\{\frac{g}{2}F_{\alpha\beta},f^{(0)} \right\}\delta'(p^2)
+ \left\{\Pi_\nu^{(1)},p^\nu f^{(0)} \right\} \delta'(p^2)\right]\nonumber\\
& &+ \frac{1}{2 p_n}\left(\left\{\bar\Pi^{(1)\mu}, p_n  f^{(0)}\delta(p^2)  \right\}
-  \left\{\Pi^{(1)}_n,  \bar p^\mu  f^{(0)}\delta(p^2)  \right\}\right)\nonumber\\
& & {+}\frac{s}{2 p_n}\bar\epsilon^{\mu\alpha\beta} \left[G_\alpha^{(0)}, \bar p_\beta f^{(0)}\delta(p^2) \right].
\end{eqnarray}

As we note in the zeroth order case, the equation (\ref{J-c2-bar-0-W}) is automatically satisfied once we have the expression (\ref{Jnbar-Jn-0}).
Now we can check if  the first order equation (\ref{J-c2-bar-1-W}) also holds automatically  by using the first order expression (\ref{Jnbar-Jn-1a})
together with Eqs.(\ref{J-t-n-0-W}), (\ref{Jn0-onshell}) and (\ref{Jmu-f-0}). In consequence, after direct calculation we find that the first order equation (\ref{J-c2-bar-1-W})
is not satisfied automatically but lead to the constraint equation for
${\mathscr{J}}_\mu^{(0)}$ or $f^{(0)}$
\begin{eqnarray}
\label{constraint-n-Jmu0}
0&=&n_\alpha\left( \left[F^{\nu\alpha} ,  {\mathscr{J}}^{(0)\mu}\right]
+\left[F^{\alpha\mu} ,  {\mathscr{J}}^{(0)\nu} \right]
+\left[F^{\mu\nu} ,  {\mathscr{J}}^{(0)\alpha} \right]\right).
\end{eqnarray}
Because $n^\alpha$ is an arbitrary auxiliary vector with normalization $n^2=1$, the constraint equation should not depend on $n^\alpha$ or this equation should hold for any $n^\alpha$.
This leads to the Lorentz covariant constraint equation
\begin{eqnarray}
\label{constraint-Jmu0}
\left[F^{\nu\alpha} ,  {\mathscr{J}}^{(0)\mu}\right]
+\left[F^{\alpha\mu} ,  {\mathscr{J}}^{(0)\nu} \right]
+\left[F^{\mu\nu} ,  {\mathscr{J}}^{(0)\alpha} \right]=0,
\end{eqnarray}
which is equivalent to
\begin{eqnarray}
\label{constraint-Jmu0-a}
\left[\tilde F^{\alpha\beta} ,  {\mathscr{J}}^{(0)}_\alpha\right]=0 \ \ \ {\textrm{with}\ \ \ \tilde F^{\alpha\beta}=\frac{1}{2}\epsilon^{\alpha\beta\mu\nu}}F_{\mu\nu}.
\end{eqnarray}
In Ref.~\cite{Ochs:1998qj}, similar  constraints for $\mathscr{F}$ and $\mathscr{S}_{\mu\nu}$ in Eq.(\ref{decomposition}) had already been obtained. Such constraints only arise
in the quantum transport theory with non-Abelian gauge field. The disentanglement theorem  of Wigner functions in Abelian gauge field given in Ref.~\cite{Gao:2018wmr} show that
all these constraint equations in Abelian cases are satisfied automatically and  holds up to any order of $\hbar$.
We also notice that the first order equation (\ref{J-c2-bar-1-W}) gives the constraint for the zeroth order Wigner function ${\mathscr{J}}^{(0)\mu}$ because the first order Wigner functions
are totally canceled due to the antisymmetry of the equation. Hence in order to get the constraint for the first order Wigner function ${\mathscr{J}}^{(1)\mu}$, we need the second order
Wigner functions and equations. The second order expression of Eq.(\ref{Js-c2}) is given by
\begin{eqnarray}
\label{J-c2-n-2-W}
 0 &=& 2 \bar p^\mu {\mathscr{J}}^{(2)}_n - 2p_n\bar{ {\mathscr{J}}}^{(2)\mu}\nonumber\\
& & + \left\{\bar\Pi^{(1)\mu}, {\mathscr{J}}^{(1)}_n \right\}    -\left\{\Pi^{(1)}_n, \bar{{\mathscr{J}}}^{(1)\mu} \right\}
 + \left\{\bar\Pi^{(2)\mu}, {\mathscr{J}}^{(0)}_n \right\}    -\left\{\Pi^{(2)}_n, \bar{{\mathscr{J}}}^{(0)\mu} \right\}\nonumber\\
& & {+} s\bar \epsilon^{\mu\alpha\beta}\left( \left[G_\alpha^{(0)}, {\mathscr{J}}_\beta^{(1)} \right]
+\left[G_\alpha^{(1)}, {\mathscr{J}}_\beta^{(0)} \right]\right),\\
\label{J-c2-bar-2-W}
0&=& 2 \bar p^\mu \bar{{\mathscr{J}}}^{(2)\nu} - 2 \bar p^\nu \bar{ {\mathscr{J}}}^{(2)\mu}\nonumber\\
& & + \left\{\bar\Pi^{(1)\mu}, \bar{{\mathscr{J}}}^{(1)\nu} \right\}    - \left\{\bar\Pi^{(1)\nu},\bar{ {\mathscr{J}}}^{(1)\mu} \right\}
 + \left\{\bar\Pi^{(2)\mu},\bar{ {\mathscr{J}}}^{(0)\nu} \right\}  - \left\{\bar\Pi^{(2)\nu}, \bar{{\mathscr{J}}}^{(0)\mu} \right\}\nonumber\\
& &  {+} s \bar\epsilon^{\mu\nu\alpha}\left( \left[G_\alpha^{(0)}, {\mathscr{J}}_n^{(1)} \right]
- \left[G_n^{(0)}, {\mathscr{J}}_\alpha^{(1)} \right]
+\left[G_\alpha^{(1)}, {\mathscr{J}}_n^{(0)} \right]
-\left[G_n^{(1)}, {\mathscr{J}}_\alpha^{(0)} \right]\right).
\end{eqnarray}
From the first equation above, we can express $\bar{ {\mathscr{J}}}^{(2)\mu}$ in terms of ${\mathscr{J}}^{(2)}_n$, ${\mathscr{J}}^{(1)}_n$ and ${\mathscr{J}}^{(0)}_n$ as
\begin{eqnarray}
\label{Jnbar-Jn-1}
\bar{ {\mathscr{J}}}^{(2)\mu}&=&\bar p^\mu \frac{{\mathscr{J}}^{(2)}_n}{p_n}
 {+}\frac{s}{2p_n} \bar\epsilon^{\mu\alpha\beta}\left( \left[G_\alpha^{(0)}, {\mathscr{J}}_\beta^{(1)} \right]
+\left[G_\alpha^{(1)}, {\mathscr{J}}_\beta^{(0)} \right]\right)\nonumber\\
& & + \frac{1}{2p_n}\left(\left\{\bar\Pi^{(1)\mu}, {\mathscr{J}}^{(1)}_n \right\}
 -\left\{\Pi^{(1)}_n, \bar{{\mathscr{J}}}^{(1)\mu} \right\} \right) \nonumber\\
& & + \frac{1}{2p_n}\left(\left\{\bar\Pi^{(2)\mu}, {\mathscr{J}}^{(0)}_n \right\}
  -\left\{\Pi^{(2)}_n, \bar{{\mathscr{J}}}^{(0)\mu} \right\}\right).
\end{eqnarray}
Similar to the first order, substituting it into Eq.(\ref{J-c2-bar-2-W}) and using Eqs. (\ref{Jnbar-Jn-1a}), (\ref{Jn1-onshell}) and (\ref{Jn1-evolve})  leads to the constraint
for $ {\mathscr{J}}^{(1)\mu}$
\begin{eqnarray}
\label{constraint-Jmu1}
\left[\tilde F_{\alpha\beta}, {\mathscr{J}}^{(1)\alpha} \right]
&=&-\frac{3}{32}\left[\left[\tilde F_{\nu\alpha}\partial^p_\beta-\tilde F_{\nu\beta}\partial^p_\alpha,F^{\nu\kappa}\partial^p_\kappa\right],
 {\mathscr{J}}^{(0)\alpha} \right].
\end{eqnarray}
As we just mentioned above, these constraints are unique for non-Abelian gauge field and absent for Abelian field. Such constraints actually originate from the fact that
the ``covariant gradient expansion'' is not completely identical to an expansion in powers of $\hbar$ for non-Abelian gauge field. One difference between non-Abelian and Abelian is the operator $G_\mu^{(0)}$.
In  the non-Abelian case, the derivative in $G_\mu^{(0)}$ is covariant derivative $D_\mu$, while in Abelian case, it is ordinary space-time derivative $\partial^x_\mu $.
When we calculate high order contribution through iterative process, we will  meet the commutator $[D_\mu,D_\nu]=ig F_{\mu\nu}/\hbar$ in non-Abelian gauge field and this term will contribute to  the lower  power order,
but for the ordinary derivative such issue will never happen in Abelian gauge field. Actually, during our calculation of (\ref{constraint-Jmu1}), we find that if we do not use the constraints for
$ {\mathscr{J}}^{(0)\alpha}$ in Eq. (\ref{constraint-Jmu0}) or (\ref{constraint-Jmu0-a}) beforehand, we will have the same term as the right side of Eq. (\ref{constraint-n-Jmu0}) but with minus sign.
This term from the second order equation  will eventually cancel  the one from the first order. Although we can not give the general proof, we expect that the third order equation of (\ref{Js-c2})
 will cancel the second order result (\ref{constraint-Jmu1}) and so on. Adding all the contributions up to any high order, the constraint equation (\ref{Js-c2}) should also be satisfied automaticaly.

\section{Decomposing covariant chiral kinetic equations  in color space}

\label{sec:color}

Up to now, the Wigner function ${{\mathscr{J}}}^{\mu}$ is  still an $N \times N$ matrix in color space. Hence it is necessary to decompose the  Wigner function into  color singlet and multiplet components:
\begin{eqnarray}
 {{\mathscr{J}}}_{\mu}(x,p)&=&{\mathscr{J}}_{\mu}^{I}(x,p) {\bf{1}} + {{\mathscr{J}}}^{a}_\mu (x,p)t^a,
\end{eqnarray}
with
\begin{eqnarray}
 {\mathscr{J}}_{\mu}^{I}(x,p) &=& \frac{1}{N} \textrm{tr}{{\mathscr{J}}}_{\mu}(x,p), \, \ \ \
{{\mathscr{J}}}^{a}_\mu (x,p)  = { 2} \textrm{tr}\, t^a {{\mathscr{J}}}_{\mu}(x,p) .
\end{eqnarray}
It should be noted that we use upper index ``$I$'' to denote singlet component.
Similarly, we can decompose the operators into the color singlet and  multiplet contributions:
\begin{eqnarray}
G_\mu^{(0)} &=&D_\mu  + G_\mu^{(0)a}t^a, \ \ \
\Pi_\mu^{(1)} = \Pi_\mu^{(1)a}t^a,\ \ \
G_\mu^{(1)} = G_\mu^{(1)a}t^a,
\end{eqnarray}
where
\begin{eqnarray}
G_\mu^{(0)a} = \frac{g}{2}F_{\mu\nu}^a \partial_p^\nu, \ \ \
\Pi_\mu^{(1)a} = \frac{i g}{4}F_{\mu\nu}^a \partial_p^\nu, \ \ \
G_\mu^{(1)a} = - \frac{i g}{8 }\left(\mathscr{D}^{ac}_\lambda F^{c}_{\mu\nu} \right)\partial_p^\lambda  \partial_p^\nu,
\end{eqnarray}
with $  {\mathscr{D}}^{ac}_\lambda = \delta^{ca}  \partial^x_\lambda  + g  f^{bca} A^b_\lambda /\hbar $.
With such decomposition, the singlet and multiplet components of  Wigner functions at the zeroth order
 can be derived from Eqs.(\ref{Jmu-f-0})
\begin{eqnarray}
\label{Jmu-0-sa}
{\mathscr{J}}^{(0){I}\mu} &=& p ^\mu   f^{(0){I}} \delta(p^2),\ \ \
{\mathscr{J}}^{(0)a\mu}  =  p ^\mu   f^{(0)a} \delta(p^2),
\end{eqnarray}
which satisfy the coupled  transport equations
\begin{eqnarray}
\label{Weq-0-s}
0 &=& \partial^x_\mu {\mathscr{J}}^{(0){I}\mu} +  \frac{1}{N} G_\mu^{(0)a}{\mathscr{J}}^{(0)a\mu},\\
\label{Weq-0-a}
0 &=&  {\mathscr{D}}^{ac}_\mu {\mathscr{J}}^{(0)c\mu}
+  2 G_\mu^{(0)a}{\mathscr{J}}^{(0){I}\mu} + d^{bca}G_\mu^{(0)b}{\mathscr{J}}^{(0)c\mu}.
\end{eqnarray}
Similarly but more complicatedly, the color decomposition  of first order Wigner functions can be derived
from Eq (\ref{Jmu-f-1})
\begin{eqnarray}
\label{Jmu-1-s-1}
{\mathscr{J}}^{(1){I}\mu} &=& p^\mu f^{(1){I}}\delta(p^2)  {-} \frac{s}{2  } \epsilon^{\mu\nu\alpha\beta} p_\nu
\frac{g}{2N}F^{a}_{\alpha\beta}f^{(0)a}\delta'(p^2)\nonumber\\
& & {+}\frac{s}{2 p_n}\bar\epsilon^{\mu\alpha\beta}   p_\beta \left( \partial^x_\alpha f^{(0){I}}
+ \frac{1}{ N} G_\alpha^{(0)a} f^{(0)a}  \right) \delta(p^2),\\
\label{Jmu-1-a-1}
{\mathscr{J}}^{(1)a\mu} &=& p^\mu f^{(1)a}\delta(p^2)  {-} s \epsilon^{\mu\nu\alpha\beta}  p_\nu
\left(\frac{g}{2}F^{a}_{\alpha\beta} f^{(0){I}}
 +\frac{1}{2} d^{bca}\frac{g}{2}F^{b}_{\alpha\beta} f^{(0)c} \right)\delta'(p^2)\nonumber\\
& & {+}\frac{s}{2p_n}\bar\epsilon^{\mu\alpha\beta}p_\beta \left( \frac{}{}
 {\mathscr{D}}^{ac}_\alpha  f^{(0)c}  + 2 G_\alpha^{(0)a}  f^{(0){I}}
+  d^{bca} G_\alpha^{(0)b} f^{(0)c} \frac{}{}\right)\delta(p^2)\nonumber\\
& &+ \frac{1}{2 p_n} i f^{bca}\left( \bar\Pi^{(1)b\mu}\left[ p_n f^{(0)c} \delta(p^2)\right]
-\Pi^{(1)b}_n\left[\bar p^\mu f^{(0)c} \delta(p^2) \right]\frac{}{}\right)\nonumber\\
& &+ i f^{bca} p^\mu \left[ \Pi_\nu^{(1)b}\left(p^\nu f^{(0)c} \right)\right] \delta'(p^2),
\end{eqnarray}
which satisfy the corresponding transport equations
\begin{eqnarray}
\label{Weq-1-s}
0 &=& \partial^x_\mu {\mathscr{J}}^{(1){I}\mu} +  \frac{1}{N} G_\mu^{(0)a}{\mathscr{J}}^{(1)a\mu},\\
\label{Weq-1-a}
0 &=&  {\mathscr{D}}^{ac}_\mu  {\mathscr{J}}^{(1)c\mu}
+  2 G_\mu^{(0)a}{\mathscr{J}}^{(1){I}\mu} + d^{bca}G_\mu^{(0)b}{\mathscr{J}}^{(1)c\mu}
+ i f^{bca}G_\mu^{(1)b}{\mathscr{J}}^{(0)c\mu}.
\end{eqnarray}
In order to attain all the results above, we have used the Eq.(\ref{Lie}) repeatedly.
We note that the singlet distribution  $f^{(0){I}}$ and multiplet distribution $f^{(0){a}}$ are totally coupled with each other
even in the zeroth order transport equation, which displays the much complexity for non-Abelian chiral kinetic equation, in comparison with
chiral kinetic equation in Abelian gauge field.

\section{Frame dependence of distribution function}
\label{sec:Lorentz}

We can regard $ f(x,p)$  as the  particle distribution function in 8-dimensional phase space and
$ f^{(0)}(x,p)$ in Eq.(\ref{Jn-0}) and $f^{(1)}(x,p)$ in Eq.(\ref{Jn1-onshell-a}) are the zeroth order and first order corrections to $f(x,p)$, respectively.
However this distribution function defined in this way depends on the auxiliary vector $n^\mu$ we choose. Since
we can identify this time-like vector $n^\mu$ as the velocity of the observer in a reference frame, the distribution
function depends on the reference frame in which we define it.  In general, the distribution function in phase space
can not be Lorentz scalar when we change the reference frame from one to another. In this section, we will derive
how these distribution functions transform in different reference frames. In order to do that, we  rewrite
the zeroth  and first order results for Wigner functions with explicit dependence on  the frame  velocity $n^\mu$ as the following:
\begin{eqnarray}
{\mathscr{J}}^{(0)\mu} &=& p ^\mu \frac{n\cdot {\mathscr{J}}^{(0)}}{ n\cdot p},\\
{\mathscr{J}}^{(1)\mu} &=& p^\mu \frac{n\cdot {\mathscr{J}}^{(1)}}{ n\cdot p}
 {+}\frac{s}{2 n\cdot p }\epsilon^{\mu\nu\alpha\beta} n_\nu \left[G_\alpha^{(0)}, {\mathscr{J}}^{(0)}_\beta \right]\nonumber\\
& &+ \frac{1}{2 n\cdot p }\left(\left\{\Pi^{(1)\mu}, n\cdot {\mathscr{J}}^{(0)}\right\}
- \left\{n\cdot \Pi^{(1)}, {\mathscr{J}}^{(0)\mu} \right\}\right).
\end{eqnarray}
Of course, we can also define the particle distribution function in another reference frame with velocity $n'$,
\begin{eqnarray}
{\mathscr{J}}^{(0)\mu} &=& p ^\mu \frac{n'\cdot {\mathscr{J}}^{(0)}}{ n'\cdot p},\\
{\mathscr{J}}^{(1)\mu} &=& p^\mu \frac{n'\cdot {\mathscr{J}}^{(1)}}{ n'\cdot p}
 {+}\frac{s}{2 n'\cdot p }\epsilon^{\mu\nu\alpha\beta} n'_\nu \left[G_\alpha^{(0)}, {\mathscr{J}}^{(0)}_\beta \right]\nonumber\\
& &+ \frac{1}{2 n'\cdot p }\left(\left\{\Pi^{(1)\mu}, n'\cdot {\mathscr{J}}^{(0)}\right\}
- \left\{n'\cdot \Pi^{(1)}, {\mathscr{J}}^{(0)\mu} \right\}\right).
\end{eqnarray}
Since ${\mathscr{J}}^{(0)\mu}$ and ${\mathscr{J}}^{(1)\mu}$ should not depend on the auxiliary vector, we will get the
modified Lorentz transformation  for ${ {\mathscr{J}}^{(0)}_n}/{ p_n}$ and ${ {\mathscr{J}}^{(1)}_n}/{ p_n}$
\begin{eqnarray}
\delta\left(\frac{n\cdot {\mathscr{J}}^{(0)}}{ n\cdot p}\right) &=&
\frac{n'\cdot {\mathscr{J}}^{(0)}}{ n'\cdot p} - \frac{n\cdot {\mathscr{J}}^{(0)}}{ n\cdot p}=0,\\
\label{rule-1}
\delta\left(\frac{n\cdot {\mathscr{J}}^{(1)}}{ n\cdot p}\right) &=&
\frac{n'\cdot {\mathscr{J}}^{(1)}}{ n'\cdot p} - \frac{n\cdot {\mathscr{J}}^{(1)}}{ n\cdot p}\nonumber\\
&=& {-}\frac{s\epsilon^{\mu\nu\alpha\beta}n_\mu n'_\nu }{2(n\cdot p) (n'\cdot p) } \left[G_\alpha^{(0)}, {\mathscr{J}}^{(0)}_\beta \right]
- \frac{\left(n_\mu n'_\nu - n_\nu n'_\mu \right)}{2 (n\cdot p)(n'\cdot p) }\left\{\Pi^{(1)\mu},  {\mathscr{J}}^{(0)\nu}\right\}.
\end{eqnarray}
We note that the zeroth order ${ {\mathscr{J}}^{(0)}_n}/{ p_n}$ does not depend on the reference frame and is Lorentz scalar while the first order
${ {\mathscr{J}}^{(1)}_n}/{ p_n}$ does have non-trivial transformation  and is not Lorentz scalar when we change from reference frame $n_\mu$ to $n'_\mu$.
The first  term of the last line in Eq.(\ref{rule-1}) is just the so-called side-jump term  and the second term is unique for non-Abelian gauge field
and absent for Abelian gauge field. We can decompose the modified Lorentz  transformation  into  color singlet and multiplet components:
\begin{eqnarray}
\delta\left(\frac{n\cdot {\mathscr{J}}^{(0){I}}}{ n\cdot p}\right)&=&0,\hspace{1cm}
\delta\left(\frac{n\cdot {\mathscr{J}}^{(0)a}}{ n\cdot p}\right)=0,\\
\delta\left(\frac{n\cdot {\mathscr{J}}^{(1){I}}}{ n\cdot p}\right)
&=& {-}\frac{s \epsilon^{\mu\nu\alpha\beta}n_\mu n'_\nu}{2(n\cdot p) (n'\cdot p) } \partial_\alpha^x {\mathscr{J}}^{(0){I}}_{\beta}
 {-}\frac{s\epsilon^{\mu\nu\alpha\beta}n_\mu n'_\nu}{2(n\cdot p) (n'\cdot p) N }
 G_\alpha^{(0)a}{\mathscr{J}}^{(0)a}_\beta ,\\
\delta\left(\frac{n\cdot {\mathscr{J}}^{(1)a}}{ n\cdot p}\right)
&=& {-}\frac{s\epsilon^{\mu\nu\alpha\beta}n_\mu n'_\nu}{2(n\cdot p) (n'\cdot p) }
\left[{\mathscr{D}}^{ac}_\alpha {\mathscr{J}}^{(0)c}_\beta
 + 2 G_\alpha^{a(0)}{\mathscr{J}}^{(0){I}}_{\beta}
+  d^{bca} G_\alpha^{(0)b}{\mathscr{J}}^{(0)c}_\beta \right] \nonumber\\
& &- \frac{n_\mu n'_\nu - n_\nu n'_\mu }{2 (n\cdot p)(n'\cdot p) } i f^{bca}\Pi^{(1)b\mu} {\mathscr{J}}^{(0)c\nu}.
\end{eqnarray}
Using Eqs. (\ref{Jmu-0-sa}), (\ref{Jmu-1-s-1}) and (\ref{Jmu-1-a-1}), we obtain
the transformation  of the singlet and multiplet distribution function of  $f^{(0)}$ and $f^{(1)}$ when we define them in
different frames, respectively,
\begin{eqnarray}
\delta(p^2)\delta f^{(0){I}}&=&0,\nonumber\\
 \delta(p^2)\delta f^{(0)a} &=& 0,\\
\label{rule-fs}
\delta(p^2)\delta f^{(1){I}}&=& {-}\delta(p^2)\frac{s\epsilon^{\mu\nu\alpha\beta}n_\mu n'_\nu p_\beta }{2(n\cdot p) (n'\cdot p) }
\left[\partial_\alpha^x f^{(0){I}} +\frac{\hbar^2}{N} G_\alpha^{(0)a}  f^{(0)a} \right],\\
\label{rule-fa}
\delta(p^2)\delta f^{(1)a}&=& {-}\delta(p^2)\frac{s\epsilon^{\mu\nu\alpha\beta}n_\mu n'_\nu p_\beta}{2(n\cdot p) (n'\cdot p) }
\left[ {\mathscr{D}}^{ac}_\alpha f^{(0)c} +2  G_\alpha^{(0)a} f^{(0){I}}
+  d^{bca}  G_\alpha^{(0)b}  f^{(0)c} \right]\nonumber\\
& &- \frac{n_\mu n'_\nu - n_\nu n'_\mu}{2 (n\cdot p)(n'\cdot p) } if^{bca}\Pi^{(1)b\mu}\left(p^\nu f^{(0)c} \delta(p^2)\right).
\end{eqnarray}
These non-trivial transformation play very important role to choose some specific solutions. They will be used to derive chiral effects in the next section.

\section{Chiral effects  in non-Abelian gauge field}
\label{sec:currents}
As we all know, chiral kinetic theory tries to  incorporate chiral anomaly, a novel and prominent quantum effect,  into  kinetic approach in a consistent way.
It can describe various chiral effects originating from chiral anomaly, such as chiral magnetic effect and chiral vortical effect.
However, as far as we know, most of work in the literature on chiral kinetic theory focused on the chiral anomaly or chiral effects induced by Abelian gauge field.
In this section,  we will demonstrate how the non-Abelian  chiral effects can arise naturally in the formalism discussed in the preceding sections.
\subsection{Non-Abelian chiral anomaly}

First of all, let us consider the non-Abelian chiral anomaly. In general, we can write the zeroth order Wigner function in free Dirac field as the following,
\begin{eqnarray}
{\mathscr{J}}^{(0)ij}_{s\mu}&=& \frac{\delta_{ij}}{4\pi^3} \left[ {\theta (p_0) } n^i_s
+{\theta (-p_0) }\left(\bar n^i_s - 1\right) \right]p_\mu \delta\left(p^2\right)
\end{eqnarray}
where we have recovered the lower chirality index $s$,   the upper scripts $i$ and  $j$ indicate the color index in fundamental representation corresponding to Eq.(\ref{density-element})
and the repeated indices here do not denote summation.
The function $n_s^i/\bar n_s^i$ represent the quark/antiquark  number density with color $i$ and chirality $s$ in phase space.
They are defined as the ensemble average of the normal-ordered number density operator and are  expected to vanish at infinity in phase space. The $-1$ term in antiparticle distribution
is vacuum or Dirac sea contribution and originate from the anticommutator of the antiparticle field in
 the definition of Wigner funciton without normal ordering.  This term plays a central role to generate the chiral anomaly as pointed out in \cite{Gao:2019zhk,Fang:2020com}.
Decomposing it  in color space gives rise to
\begin{eqnarray}
{\mathscr{J}}^{(0)ij}_{s\mu}&=& \delta_{ij} {\mathscr{J}}^{(0){I}}_{s\mu}  + t^a_{ij} {\mathscr{J}}^{(0)a}_{s\mu}
\end{eqnarray}
where the singlet and multiplet components are given by, respectively,
\begin{eqnarray}
\label{J-s-0-J-a-0}
 {\mathscr{J}}^{(0){I}}_{s\mu} &=& p_\mu f^{(0){I}}_s \delta(p^2), \ \ \ \   {\mathscr{J}}^{(0)a}_{s\mu} =  p_\mu f^{(0)a}_s \delta(p^2),
\end{eqnarray}
with
\begin{eqnarray}
\label{f0ss}
 f^{(0){I}}_s&=&  \frac{1}{4\pi^3 N}\sum_i \left[{\theta (p_0) } n_s^i +{\theta (-p_0) }\bar n^i_s\right]
- \frac{1}{4\pi^3} \theta(-p_0),\\
\label{f0as}
 f^{(0)a}_s&=&  \frac{1}{2\pi^3 }\sum_i t^a_{ii}
\left[{\theta (p_0) }n^i_s + {\theta (-p_0) } \bar n^i_s \right].
\end{eqnarray}
We note that only the singlet component $f^{(0){I}}_s$ includes the  vacuum contribution. In order to
consider the chiral anomaly, we need the transport equation for the axial Wigner functions ${\mathscr{A}}^{{I}\mu}$ and ${\mathscr{A}}^{{a}\mu}$
\begin{eqnarray}
\label{Amu}
 {\mathscr{A}}^{I\mu} = \sum_{s=\pm 1} s{\mathscr{J}}^{I\mu}_{s},\hspace{2cm}
  {\mathscr{A}}^{a\mu} = \sum_{s=\pm 1} s{\mathscr{J}}^{a\mu}_{s},
\end{eqnarray}
from which we can obtain the chiral currents
\begin{eqnarray}
\label{Amu}
j^{I\mu}_5 = \int d^4 p  {\mathscr{A}}^{I\mu},\hspace{2cm}   j^{a\mu}_5 = \int d^4 p  {\mathscr{A}}^{a\mu}.
\end{eqnarray}
The zeroth order equations can be derived trivially from Eqs.(\ref{Weq-0-s},\ref{Weq-0-a})
\begin{eqnarray}
\label{A-0-s}
\partial^x_\mu {\mathscr{A}}^{(0){I}\mu} &=&  -  \frac{1}{N} G_\mu^{(0)a}{\mathscr{A}}^{(0)a\mu},\\
\label{A-0-a}
{\mathscr{D}}^{ac}_\mu {\mathscr{A}}^{(0)c\mu} &=&
-  2 G_\mu^{(0)a}{\mathscr{A}}^{(0){I}\mu} - d^{bca}G_\mu^{(0)b}{\mathscr{A}}^{(0)c\mu}.
\end{eqnarray}
From the expression (\ref{J-s-0-J-a-0}), we note that the  vacuum contributions in $ {\mathscr{A}}^{(0){I}\mu}$ and $ {\mathscr{A}}^{(0){a}\mu}$ are all cancelled between $s=+1$ and $s=-1$. Since the right hand sides of the  equations above are all total derivatives on momentum and
only normal particle distributions are involved, integrating over the 4-momentum  leads to the conservation of chiral current at the zeroth order.
\begin{eqnarray}
\partial^x_\mu j^{(0){I}\mu}_5  =0,\hspace{1cm}
{\mathscr{D}}^{ac}_\mu  j^{(0)c\mu}_5 =0.
\end{eqnarray}
The first order equations can be given  from Eqs.(\ref{Weq-1-s}, \ref{Weq-1-a})
\begin{eqnarray}
\label{A-1-s}
0 &=& \partial^x_\mu {\mathscr{A}}^{(1){I}\mu} +  \frac{1}{N} G_\mu^{(0)a}{\mathscr{A}}^{(1)a\mu},\\
\label{A-1-a}
0 &=&  {\mathscr{D}}^{ac}_\mu  {\mathscr{A}}^{(1)c\mu}
+  2 G_\mu^{(0)a}{\mathscr{A}}^{(1){I}\mu} + d^{bca}G_\mu^{(0)b}{\mathscr{A}}^{(1)c\mu}
+ i f^{bca}G_\mu^{(1)b}{\mathscr{A}}^{(0)c\mu}.
\end{eqnarray}
The right hand sides of these first order equations are still all total derivatives, after integrating over momentum, the only possible
nonvanishing contribution is from the singular vacuum term,
\begin{eqnarray}
\partial^x_\mu j^{(1){I}\mu}_5  &=&
 \frac{ g^2 }{2N} F^a_{\mu\lambda}\tilde F^{a,\mu\nu}
\int d^4 p\, \partial^\lambda_p\left[ p_\nu f_v^{(0)} \delta'(p^2)\right],\\
{\mathscr{D}}^{ac}_\mu  j^{(1)c\mu}_5 &=&
  \frac{g^2}{2} d^{bca}F^b_{\mu\lambda}\tilde F^{c,\mu\nu}
\int d^4 p\, \partial^\lambda_p \left[ p_\nu f_v^{(0)}\delta'(p^2)\right],
\end{eqnarray}
where $f_v^{(0)}$ represents the vacuum  contribution
\begin{eqnarray}
 f_v^{(0)}&=& - \frac{1}{2\pi^3} \theta(-p_0).
\end{eqnarray}
Using the identity
\begin{eqnarray}
F^a_{\mu\lambda} \tilde F^{a,\mu\nu}  &=& \frac{1}{4}g^\nu_\lambda\tilde F^{\alpha\beta} F_{\alpha\beta}
=g^\nu_\lambda E^a\cdot B^a,\\
d^{bca}F^b_{\mu\lambda} \tilde F^{c,\mu\nu}  &=& \frac{1}{4}g^\nu_\lambda d^{bca}\tilde F^{b,\alpha\beta} F^c_{\alpha\beta}
=g^\nu_\lambda d^{bca} E^b\cdot B^c,
\end{eqnarray}
we have
\begin{eqnarray}
\partial^x_\mu j^{(1){I}\mu}_5  &=&
\frac{ g^2}{2 N} E^a\cdot B^a
\int d^4 p\, \partial^\lambda_p\left[ p_\lambda f_v^{(0)} \delta'(p^2)\right],\\
{\mathscr{D}}^{ac}_\mu  j^{(1)c,\mu}_5 &=&
\frac{g^2}{2} d^{bca}E^b\cdot B^c
\int d^4 p\, \partial^\lambda_p \left[ p_\lambda f_v^{(0)}\delta'(p^2)\right].
\end{eqnarray}
As in the Abelian case \cite{Gao:2019zhk,Fang:2020com} , we can finish integrating the momentum
\begin{eqnarray}
C_v=\int d^4 p\, \partial^\lambda_p \left[ p_\lambda f_v^{(0)}\delta'(p^2)\right]
\end{eqnarray}
in 4 dimensional Euclidean momentum space $p_E^\mu = (ip_0, {\bf p})$ by Wick rotation
\begin{eqnarray}
C_v = -\frac{1}{2\pi^2}\int \frac{d^{4}p_E}{2\pi^2}\, \partial_{\mu}\left(\frac{p^{\mu}_E}{p_E^4}\right)=-\frac{1}{2\pi^2},
\end{eqnarray}
or 3 dimensional Euclidean momentum space ${\bf p}$ after integrating over $p_0$
\begin{eqnarray}
C_v=-\frac{1}{2\pi^2}\int \frac{d^3{\bf p}}{2\pi}{\mathbf \partial}_{\bf p}\cdot \left( \,
\frac{\hat{\bf p}}{2{\bf p}^2}\right)
=-\frac{1}{2\pi^2},
\end{eqnarray}
where $p^\mu_E/p^4_E$ and $\hat{\bf p}/2{\bf p}^2$ are just the Berry curvature of a 4-dimensional and 3-dimensional monopoles in Euclidean momentum space, respectively.
It follows that
\begin{eqnarray}
\partial^x_\mu j^{(1){I}\mu}_5 &=&
 {-}\frac{ g^2 }{4\pi^2 N} E^a\cdot B^a,\ \ \
{\mathscr{D}}^{ac}_\mu  j^{(1)c\mu}_5 =
 {-}\frac{g^2}{4\pi^2 } d^{bca}E^b\cdot B^c.
\end{eqnarray}
It is obvious that the non-Abelian chiral anomaly  originates from   the Berry curvature of the vacuum contribution.

\subsection{Non-Abelian anomalous currents}

As we all know that the vorticity and magnetic  field imposed on a chiral system could induce some novel chiral effects such as chiral magnetic effect, chiral vortical effect and chiral separate effect.
In this section, we will derive the chiral effects induced by non-Abelian gauge field. For the  zeroth order distribution function in Eqs.(\ref{f0ss},\ref{f0as}), we assume the quark and antiquark
number density is the global equilibrium Fermi-Dirac distribution
\begin{eqnarray}
\label{n-FD}
n_s^i = \frac{1}{1+ e^{ \left(u\cdot p-\mu_s^i \right)/T }}, \ \ \ \
\bar n_s^i = \frac{1}{1+ e^{ \left(-u\cdot p + \mu_s^i \right)/T }}.
\end{eqnarray}
where $\mu_s^i$ denotes the chemical potential of the quark with chirality $s$ and color $i$. The chirality chemical potential $\mu_s^i$ is related to the vector chemical potential $\mu^i$
and axial chemical potential $\mu_5^i$ by  $ \mu_s^i =  \mu^i  {+} s \mu_5^i$.
Now let us  impose  the covariant-constant field  in this chiral system
\begin{eqnarray}
F^a_{\mu\nu}= F_{\mu\nu}\xi^a
\end{eqnarray}
with  the color index $a$ only running in the $N-1$ commuting Cartan generators and  $\xi^a$ being ($N-1$) - dimensional constant color vector. Since the field tensor
$F_{\mu\nu}$  is independent of space and time,   the external gauge potential $A^a_\mu$
can be  chosen as
\begin{eqnarray}
A^a_{\mu}=-\frac{1}{2} F_{\mu\nu}x^\nu \xi^a.
\end{eqnarray}
It is easy to verify that when the following constraint conditions are satisfied
\begin{eqnarray}
\label{conditions}
\partial^x_\mu \frac{u_\nu}{T} + \partial^x_\nu \frac{u_\mu}{T}=0,\ \ \  \partial_\mu^x \frac{\mu^i_s}{T} =g\xi^a t^a_{ii}\frac{ E_\mu}{T},
\end{eqnarray}
the zeroth order Wigner function in (\ref{f0ss},\ref{f0as}) with Fermi-Dirac distribution is indeed the solution of the zeroth order Wigner equations (\ref{Weq-0-s}, \ref{Weq-0-a}).
Once we have a special zeroth order solution, most of the terms in the first order solution are totally fixed by Eqs.(\ref{Jmu-1-s-1}) and (\ref{Jmu-1-a-1}) except for
the first terms with $f^{(1)I}_s=0$ or $f^{(1)a}_s=0$.  As shown in Ref.~\cite{Gao:2018jsi},  we can not causally set $f^{(1)I}_s=0$ and $f^{(1)a}_s=0$ because they
must be consistent with the transformations  (\ref{rule-fs}) and (\ref{rule-fa}). Substituting these specific solution (\ref{f0ss},\ref{f0as},\ref{n-FD}) and conditions (\ref{conditions})
into the transformations of the first order, we can have
\begin{eqnarray}
\label{rule-fs-eq}
\delta(p^2)\delta f_s^{(1){I}}&=& {-}\delta(p^2)\frac{s n'_\nu \tilde \Omega^{\nu\sigma}p_\sigma }{2(n'\cdot p) }\frac{d f_s^{(0){I}}}{d y}
 {+}\delta(p^2)\frac{s n_\nu \tilde \Omega^{\nu\sigma}p_\sigma }{2(n\cdot p) }\frac{d f_s^{(0){I}}}{d y},\\
\label{rule-fa-eq}
\delta(p^2)\delta f_s^{(1)a}&=& {-}\delta(p^2)\frac{s n'_\nu \tilde \Omega^{\nu\sigma}p_\sigma }{2(n'\cdot p) }\frac{d f_s^{(0)a}}{d y}
 {+}\delta(p^2)\frac{s n_\nu \tilde \Omega^{\nu\sigma}p_\sigma }{2(n\cdot p) }\frac{d f_s^{(0)a}}{d y},
\end{eqnarray}
where we have defined
\begin{eqnarray}
\Omega_{\mu\nu}&=&\frac{1}{2}\left(\partial^x_\mu \frac{u_\nu}{T} - \partial^x_\nu \frac{u_\mu}{T} \right),\ \ \
\tilde\Omega_{\mu\nu} = \frac{1}{2}\epsilon_{\mu\nu\alpha\beta}\Omega^{\alpha\beta},\ \ \
y =  u\cdot p/T.
\end{eqnarray}
This   indicates that we can choose the specific solution which is consistent with the transformations (\ref{rule-fs}) and (\ref{rule-fa}),
\begin{eqnarray}
 f^{(1){I}}_s &=& {-}\frac{s n_\nu \tilde \Omega^{\nu\sigma}p_\sigma }{2(n\cdot p) }\frac{d f^{(0){I}}_s}{d y},\ \ \
 f^{(1)a}_s= {-}\frac{s n_\nu \tilde \Omega^{\nu\sigma}p_\sigma }{2(n\cdot p) }\frac{d f^{(0)a}_s}{d y}.
\end{eqnarray}
Inserting  these results into Eqs.(\ref{Jmu-1-s-1}) and (\ref{Jmu-1-a-1}) gives rise to
\begin{eqnarray}
\label{Jmu-1-s-FD}
{\mathscr{J}}_s^{(1)I\mu} &=& {-} \frac{s}{2}\tilde\Omega^{\mu\nu}p_\nu \frac{d f^{(0)I}_s}{d y} \delta(p^2)
  {-} \frac{s g }{2 N } \tilde F^{a,\mu\nu} p_\nu f^{(0)a}_s\delta'(p^2),\\
\label{Jmu-1-a-FD}
{\mathscr{J}}^{(1)a,\mu}_s &=& {-}\frac{s}{2}\tilde\Omega^{\mu\nu}p_\nu \frac{d f^{(0)a}_s}{d y} \delta(p^2)
  {-} {s g} \tilde F^{b,\mu\nu}  p_\nu \left(\delta^{ab} f^{(0)I}_s
 +\frac{1}{2} d^{bca} f^{(0)c}_s \right)\delta'(p^2),
\end{eqnarray}
where we have dropped all the terms which vanish when color index  runs only in  the $N-1$ commuting Cartan generators. It is obvious that the final expressions do not depend on
the auxiliary vector $n^\mu$ any more and are explicitly  Lorentz covariant.

Now it is straightforward to obtain the right-handed/left-handed  currents  by integrating  the 4-dimension momentum $p^\mu$.
\begin{eqnarray}
j_s^{(1)I\mu} &=& {-}\frac{s}{2}\omega^\mu  \int d^4 p \, y \frac{d f^{(0)I}_s}{d y} \delta(p^2)
 {-} \frac{s g }{2N } B^{a\mu} T \int d^4 p\,  y f^{(0)a}_s\delta'(p^2),\\
j^{(1)a\mu}_s &=& {-} \frac{s}{2}\omega^\mu  \int d^4 p \, y \frac{d f^{(0)a}_s}{d y} \delta(p^2)
  {-} {s g} B^{b\mu} T \int d^4 p\, y (\delta^{ab} f^{(0)I}_s
 +\frac{ d^{bca}}{2} f^{(0)c}_s )\delta'(p^2),
\end{eqnarray}
where $\omega^\mu =T \tilde\Omega^{\mu\nu}u_\nu =\epsilon^{\mu\nu\alpha\beta}u_\nu \partial_\alpha^x u_\beta/2$.
From Eqs.(\ref{f0ss}) and (\ref{f0as}) together with Eq.(\ref{n-FD}), we can finish  the integrals analytically
\begin{eqnarray}
 \int d^4 p\, y \frac{d f^{(0)I}_s}{dy}\delta(p^2) &=& -\frac{T^2}{6} - \frac{ \sum_i \mu_s^{i\,2}}{2\pi^2 N },\ \ \ \
T \int d^4 p \, y f^{(0)I}_s\delta'(p^2) = \frac{\sum_i \mu_s^i}{4\pi^2 N} ,\\
\int d^4 p\, y \frac{df^{(0)a}_s}{dy}\delta(p^2)&=&-\frac{\sum_i t^a_{ii}\mu_s^{i\,2}}{\pi^2 } ,\ \ \ \ \ \ \ \ \
 T \int d^4 p \, y  f^{(0)a}_s\delta'(p^2)=\frac{\sum_i t^a_{ii}\mu_s^i}{2\pi^2 } ,\
\end{eqnarray}
It follows that
\begin{eqnarray}
j_s^{(1)I\mu} &=&\xi_s^{I} \omega^{\mu} +  \xi_{Bs}^{Ia} B^{a\mu} ,\ \ \
j^{(1)a\mu}_s = \xi_s^a \omega^{\mu}   +   \xi_{Bs}^{ab}B^{b\mu}
\end{eqnarray}
where
\begin{eqnarray}
\xi_s^{I} &=&  s \left(\frac{T^2}{12} + \frac{1}{4\pi^2 N }  \sum_i \mu_s^{i\,2}\right),\ \ \ \
\xi_{Bs}^{Ia} =  {-}   \frac{s g }{4\pi^2  N } \sum_i t^a_{ii}\mu_s^i  ,\\
\xi_s^a &=&   \frac{s}{2\pi^2  }  \sum_i t^a_{ii} \mu_s^{i\,2},\hspace{2cm}
 \xi_{Bs}^{ab} =  {-}\frac{s g}{4\pi^2 }  \left( \frac{\delta^{ab}}{ N} \sum_i \mu_s^i
 + {d^{bca}}  \sum_i t^c_{ii} \mu_s^i \right).
\end{eqnarray}
The vector current and axial current can be obtained from right-hand and left-hand currents directly,
\begin{eqnarray}
j^{(1)I\mu} &=& j^{(1)I\mu}_{+1} + j^{(1)I\mu}_{-1} = \xi^{I} \omega^{\mu} +  \xi_{B}^{Ia} B^{a\mu} ,\\
j^{(1)a\mu} &=&j^{(1)a\mu}_{+1} + j^{(1)a\mu}_{-1} = \xi^a \omega^{\mu}   +   \xi_{B}^{ab}B^{b\mu},\\
j^{(1)I\mu}_5 &=& j^{(1)I\mu}_{+1} - j^{(1)I\mu}_{-1} = \xi^{I}_5 \omega^{\mu} +  \xi_{B5}^{Ia} B^{a\mu} ,\\
j^{(1)a\mu}_5 &=&j^{(1)a\mu}_{+1} - j^{(1)a\mu}_{-1} = \xi^a_5 \omega^{\mu}   +   \xi_{B5}^{ab}B^{b\mu},
\end{eqnarray}
where the anomalous transport coefficients for the vector currents are given by
\begin{eqnarray}
\xi^{I} &=& \frac{1}{\pi^2 N }  \sum_i \mu^{i}\mu_5^i,\ \ \ \
\xi_{B}^{Ia} =  -  \frac{ g }{2\pi^2 N } \sum_i t^a_{ii}\mu_5^i  ,\\
\xi^a &=&   \frac{2}{\pi^2  }  \sum_i t^a_{ii} \mu^{i}\mu_5^i,\ \ \
 \xi_{B}^{ab} = - \frac{ g}{2\pi^2 }  \left( \frac{\delta^{ab}}{ N} \sum_i \mu_5^i
 + {d^{bca}}\sum_i t^c_{ii} \mu_5^i \right)
\end{eqnarray}
and the coefficients for the axial currents are given by
\begin{eqnarray}
\xi_5^{I} &=&  \frac{T^2}{6}  {+} \frac{1}{2\pi^2 N }  \sum_i (\mu^{i\,2}+\mu_5^{i\,2}),\ \
\xi_{B5}^{Ia} =  {-}   \frac{ g }{2\pi^2  N } \sum_i t^a_{ii}\mu^i  ,\\
\xi_5^a &=&     \frac{1}{\pi^2 }  \sum_i t^a_{ii}( \mu^{i\,2} +  \mu_5^{i\,2}),\hspace{1cm}
 \xi_{B5}^{ab} =  {-}\frac{ g}{2\pi^2 }  \left( \frac{\delta^{ab}}{ N} \sum_i \mu^i
 + {d^{bca}} \sum_i t^c_{ii} \mu^i \right).
\end{eqnarray}
These are just the non-Abelian counterparts of the  chiral magnetic effect, chiral vortical effect and chiral separation  effect.
We note that the coefficients $\xi^I$, $\xi_B^{Ia}$, $\xi^I_5$ and  $\xi_{B 5}^{Ia}$ for the singlet current are very similar to the coefficients in the Abelian case. They can be regarded
as the average value of the coefficient in Abelian currents over different colors. These results will reduce into the usual Abelian chiral effects if  we set $N=1$, $t^a_{ii}=1$ and $g=-1$.
The coefficients $\xi^a$, $\xi_B^{ab}$, $\xi^a_5$ and  $\xi_{B 5}^{ab}$ are unique for the non-Abelian currents and similar results were also  obtained in different approachs in Refs.
\cite{Son:2009tf,Landsteiner:2011cp}.

\section{Summary}

\label{sec:summary}

In this paper, we generalize the chiral kinetic theory in Abelian gauge field to non-Abelian gauge field. Starting from  the gauge invariant and Lorentz invariant quantum transport theory
set up in \cite{Heinz:1983nx,Elze:1986hq,Elze:1986qd,Elze:1989un,Ochs:1998qj}, we  decompose  the Wigner functions and Wigner equations completely both in spinor space and  in color space.
With the help of the ``covariant gradient expansion'',
we find that the right-handed and left-handed Wigner function are totally decoupled with all the other Wigner functions. Among the four components of right-handed or left-handed Wigner functions,
we can define the time-like component as the independent Wigner function and regard it as the phase space particle distribution function in some reference frame with velocity $n^\mu$.
In consequence, all the space-like components can be totally determined by this chosen independent distribution function. Such disentangling process simplifies the Wigner equations greatly.
The difference between Abelian  and non-Abelian gauge field is that in Abelian gauge field  the disentanglement theorem demonstrated in \cite{Gao:2018wmr} show that the transport equation
for space-like components are automatically satisfied while in non-Abelian gauge field these equations are not satisfied automatically order by order and we obtain extra constraint conditions.
We  present the  chiral kinetic equations up to the first order in non-Abelian gauge field  in 8-dimension phase space. Since the  kinetic equations of the singlet component and multiplet components are totally
coupled with each other, the non-Abelian chiral kinetic equation is much more complicated than Abelian chiral kinetic equation.  We also give
the modified Lorentz transformation of the non-Abelian phase space distribution function when we define them in different frames.  Finally, we  utilize it to
calculate the non-Abelian chiral anomaly and the vector and axial currents induced by color field and vorticity and and find that it is  consistent and successful
in describing the chiral effects in non-Abelian gauge field.


\acknowledgments

This work was supported in part by  the National Natural
Science Foundation of China  under Grant
Nos. 11890710, 11890713 and 11475104, and
the Natural Science Foundation of Shandong Province under Grant
No. JQ201601.





\begin{thebibliography}{99}






\bibitem{Vilenkin:1980fu}
  A.~Vilenkin,
  { Phys.\ Rev.\ D} {\bf 22}, 3080 (1980).

\bibitem{Kharzeev:2007jp}
  D.~E.~Kharzeev, L.~D.~McLerran and H.~J.~Warringa,
  { Nucl.\ Phys.\ A} {\bf 803}, 227 (2008).

\bibitem{Fukushima:2008xe}
  K.~Fukushima, D.~E.~Kharzeev and H.~J.~Warringa,
  { Phys.\ Rev.}\ D {\bf 78}, 074033 (2008).


\bibitem{Vilenkin:1978hb}
  A.~Vilenkin,
  {Phys.\ Lett.}\  {\bf 80B}, 150 (1978).


\bibitem{Kharzeev:2007tn}
  D.~Kharzeev and A.~Zhitnitsky,
  {Nucl.\ Phys.\ A} {\bf 797} , 67(2007).

\bibitem{Erdmenger:2008rm}
  J.~Erdmenger, M.~Haack, M.~Kaminski and A.~Yarom,
  {JHEP} {\bf 0901}, 055 (2009).

\bibitem{Banerjee:2008th}
  N.~Banerjee, J.~Bhattacharya, S.~Bhattacharyya, S.~Dutta, R.~Loganayagam and P.~Surowka,
  {JHEP} {\bf 1101}, 094 (2011).



\bibitem{Son:2004tq}
  D.~T.~Son and A.~R.~Zhitnitsky,
  {Phys.\ Rev.\ D} {\bf 70}, 074018(2004).

\bibitem{Metlitski:2005pr}
  M.~A.~Metlitski and A.~R.~Zhitnitsky,
  {Phys.\ Rev.\ D} {\bf 72}, 045011 (2005).



\bibitem{Kharzeev:2013ffa}
D.~E.~Kharzeev,
{Prog. Part. Nucl. Phys.} \textbf{75}, 133-151 (2014).


\bibitem{Kharzeev:2015znc}
D.~E.~Kharzeev, J.~Liao, S.~A.~Voloshin and G.~Wang,
{Prog. Part. Nucl. Phys.} \textbf{88}, 1-28 (2016).

\bibitem{Liu:2020ymh}
Y.~C.~Liu and X.~G.~Huang,
{ Nucl. Sci. Tech.} \textbf{31}, no.6, 56 (2020).


\bibitem{Gao:2020vbh}
J.~H.~Gao, G.~L.~Ma, S.~Pu and Q.~Wang,
Nucl. Sci. Tech. \textbf{31}, no.9, 90 (2020).

\bibitem{Gao:2020pfu}
J.~H.~Gao, Z.~T.~Liang and Q.~Wang,
Int. J. Mod. Phys. A \textbf{36}, no.01, 2130001 (2021)



\bibitem{Duval:2005vn}
  C.~Duval, Z.~Horvath, P.~A.~Horvathy, L.~Martina and P.~Stichel,
  Mod.\ Phys.\ Lett.\ B {\bf 20} (2006) 373

\bibitem{Wong:2011nt}
  C.~H.~Wong and Y.~Tserkovnyak,
  Phys.\ Rev.\ B {\bf 84} (2011) 115209

\bibitem{Son:2012wh}
  D.~T.~Son and N.~Yamamoto,
  Phys.\ Rev.\ Lett.\  {\bf 109} (2012) 181602

\bibitem{Stephanov:2012ki}
  M.~A.~Stephanov and Y.~Yin,
  Phys.\ Rev.\ Lett.\  {\bf 109} (2012) 162001

\bibitem{Stone:2013sga}
  M.~Stone and V.~Dwivedi,
  Phys.\ Rev.\ D {\bf 88} (2013) no.4,  045012


\bibitem{Dwivedi:2013dea}
  V.~Dwivedi and M.~Stone,
  J.\ Phys.\ A {\bf 47} (2013) 025401

\bibitem{Akamatsu:2014yza}
  Y.~Akamatsu and N.~Yamamoto,
  Phys.\ Rev.\ D {\bf 90} (2014) no.12,  125031


\bibitem{Chen:2014cla}
  J.~Y.~Chen, D.~T.~Son, M.~A.~Stephanov, H.~U.~Yee and Y.~Yin,
  Phys.\ Rev.\ Lett.\  {\bf 113} (2014) no.18,  182302

\bibitem{Manuel:2014dza}
  C.~Manuel and J.~M.~Torres-Rincon,
  Phys.\ Rev.\ D {\bf 90} (2014) no.7,  076007

\bibitem{Hayata:2017ihy}
  T.~Hayata and Y.~Hidaka,
  PTEP {\bf 2017} (2017) no.7,  073I01


\bibitem{Gao:2012ix}
  J.~H.~Gao, Z.~T.~Liang, S.~Pu, Q.~Wang and X.~N.~Wang,
  Phys.\ Rev.\ Lett.\  {\bf 109}, 232301 (2012)




\bibitem{Chen:2012ca}
  J.~W.~Chen, S.~Pu, Q.~Wang and X.~N.~Wang,
  Phys.\ Rev.\ Lett.\  {\bf 110} (2013) no.26,  262301






\bibitem{Hidaka:2016yjf}
  Y.~Hidaka, S.~Pu and D.~L.~Yang,
  Phys.\ Rev.\ D {\bf 95} (2017) no.9,  091901




\bibitem{Huang:2018wdl}
  A.~Huang, S.~Shi, Y.~Jiang, J.~Liao and P.~Zhuang,
  Phys.\ Rev.\ D {\bf 98} (2018) no.3,  036010


\bibitem{Gao:2018wmr}
  J.~H.~Gao, Z.~T.~Liang, Q.~Wang and X.~N.~Wang,
  Phys.\ Rev.\ D {\bf 98} (2018) no.3,  036019



\bibitem{Liu:2018xip}
  Y.~C.~Liu, L.~L.~Gao, K.~Mameda and X.~G.~Huang,
  Phys.\ Rev.\ D {\bf 99} (2019) no.8,  085014


\bibitem{Son:2012zy}
  D.~T.~Son and N.~Yamamoto,
  Phys.\ Rev.\ D {\bf 87} (2013) 085016

\bibitem{Carignano:2018gqt}
  S.~Carignano, C.~Manuel and J.~M.~Torres-Rincon,
  Phys.\ Rev.\ D {\bf 98} (2018) no.7,  076005

\bibitem{Lin:2019ytz}
  S.~Lin and A.~Shukla,
  JHEP {\bf 1906} (2019) 060


\bibitem{Carignano:2019zsh}
  S.~Carignano, C.~Manuel and J.~M.~Torres-Rincon,
  arXiv:1908.00561 [hep-ph].


\bibitem{Mueller:2017lzw}
  N.~Mueller and R.~Venugopalan,
  Phys.\ Rev.\ D {\bf 97} (2018) no.5,  051901

\bibitem{Mueller:2017arw}
  N.~Mueller and R.~Venugopalan,
  Phys.\ Rev.\ D {\bf 96} (2017) no.1,  016023

\bibitem{Mueller:2019gjj}
  N.~Mueller and R.~Venugopalan,
  Phys.\ Rev.\ D {\bf 99} (2019) no.5,  056003

\bibitem{Sun:2016nig}
  Y.~Sun, C.~M.~Ko and F.~Li,
  Phys.\ Rev.\ C {\bf 94} (2016) no.4,  045204

\bibitem{Sun:2016mvh}
  Y.~Sun and C.~M.~Ko,
  Phys.\ Rev.\ C {\bf 95} (2017) no.3,  034909

\bibitem{Sun:2017xhx}
  Y.~Sun and C.~M.~Ko,
  Phys.\ Rev.\ C {\bf 96} (2017) no.2,  024906


\bibitem{Sun:2018idn}
  Y.~Sun and C.~M.~Ko,
  Phys.\ Rev.\ C {\bf 98} (2018) no.1,  014911

\bibitem{Sun:2018bjl}
  Y.~Sun and C.~M.~Ko,
  Phys.\ Rev.\ C {\bf 99} (2019) no.1,  011903


\bibitem{Zhou:2018rkh}
  W.~H.~Zhou and J.~Xu,
  Phys.\ Rev.\ C {\bf 98} (2018) no.4,  044904


\bibitem{Zhou:2019jag}
  W.~H.~Zhou and J.~Xu,
  Phys.\ Lett.\ B {\bf 798} (2019) 134932

\bibitem{Liu:2019krs}
  S.~Y.~F.~Liu, Y.~Sun and C.~M.~Ko,
  Phys. Rev. Lett. \textbf{125}, no.6, 062301 (2020)




\bibitem{Gribov:1984tu}
  L.~V.~Gribov, E.~M.~Levin and M.~G.~Ryskin,
  Phys.\ Rept.\  {\bf 100} (1983) 1.

\bibitem{Mueller:1985wy}
  A.~H.~Mueller and J.~w.~Qiu,
  Nucl.\ Phys.\ B {\bf 268} (1986) 427.

\bibitem{McLerran:1993ka}
  L.~D.~McLerran and R.~Venugopalan,
  Phys.\ Rev.\ D {\bf 49} (1994) 3352
\bibitem{McLerran:1994vd}
  L.~D.~McLerran and R.~Venugopalan,
  Phys.\ Rev.\ D {\bf 50} (1994) 2225

\bibitem{Iancu:2003xm}
  E.~Iancu and R.~Venugopalan,
  In *Hwa, R.C. (ed.) et al.: Quark gluon plasma* 249-3363



\bibitem{Heinz:1983nx}
  U.~W.~Heinz,
  Phys.\ Rev.\ Lett.\  {\bf 51}, 351 (1983).



\bibitem{Elze:1986hq}
  H.~T.~Elze, M.~Gyulassy and D.~Vasak,
  Phys.\ Lett.\ B {\bf 177}, 402 (1986).

\bibitem{Elze:1986qd}
  H.~T.~Elze, M.~Gyulassy and D.~Vasak,
  Nucl.\ Phys.\  B {\bf 276}, 706 (1986).



\bibitem{Elze:1989un}
  H.~T.~Elze and U.~W.~Heinz,
  Phys.\ Rept.\  {\bf 183}, 81 (1989).




\bibitem{Groot:1980}
  S.R. De Groot, W.A. Van Leeuwen, and C.G. Van Weert, ``Relativistic Kinetic Theory''
   North-Holland, Amsterdam, 1980

\bibitem{Gao:2019zhk}
  J.~H.~Gao, Z.~T.~Liang and Q.~Wang,
  Phys. Rev. D \textbf{101}, no.9, 096015 (2020)

\bibitem{Fang:2020com}
R.~H.~Fang and J.~H.~Gao,
Nucl. Phys. A \textbf{1005}, 121851 (2021)


\bibitem{Ochs:1998qj}
  S.~Ochs and U.~W.~Heinz,
  Annals Phys.\  {\bf 266}, 351 (1998)

\bibitem{Itzykson}
  C. Itzykson and J. Zuber,
  ``Quantum Field Thoery''
  McGraw-Hill International Book Company.


\bibitem{Gao:2018jsi}
J.~h.~Gao, J.~Y.~Pang and Q.~Wang,
Phys. Rev. D \textbf{100}, no.1, 016008 (2019)

\bibitem{Son:2009tf}
  D.~T.~Son and P.~Surowka,
  { Phys.\ Rev.\ Lett.}\  {\bf 103}, 191601 (2009).



\bibitem{Landsteiner:2011cp}
  K.~Landsteiner, E.~Megias and F.~Pena-Benitez,
  { Phys.\ Rev.\ Lett.}\  {\bf 107}, 021601 (2011).



\end{thebibliography}
\end{document}